\documentclass[10pt,a4,twocolumn]{IEEEtran}
\usepackage{epsfig,fullpage,times,url,graphicx}

\author{SeonYeong Han and Nael B. Abu-Ghazaleh\\
Computer Science Dept.\\
State University of New York at Binghamton\\
\url{{shan6@,nael@cs.}binghamton.edu}}

\title{On the Effect of Fading on Ad-hoc Networks}
\begin{document}
\pagestyle{empty}
				  
\maketitle

\begin{abstract}~\label{abstract}

Most MANET (Mobile Ad hoc NETwork) research assumes idealized
propagation models.  Experimental results have shown significant
divergence from simulation results due to the effect of signal fading
in realistic wireless communication channels.  In this paper, we
characterize the impact of fading on protocol performance. We first
study the effect of fading on MAC performance and show that its effect
can be dominating.  One of our important conclusions is that
eliminating RTS/CTS packets results in more effective operation under
fading.  We also identify an unfairness problem that arises due to
backoffs in the presence of fading.  Moreover, fading results in
several subtle interactions between the MAC and routing layers.  We
identify several of these problems and make observations about
effective approaches for addressing them.  For example, the criteria
for determining the best path should not only consider the link status
but also the link order. In addition, because routing protocols rely
on MAC level transmission failure (when the retry limit is exceeded),
route failure errors are often generated unnecessarily. Finally,
because MAC level broadcasts are unreliable, they are especially
vulnerable to fading. We analyze these effects and outline preliminary
solutions to them.
\end{abstract}

\section{Introduction}

Mobile Ad hoc NETworks (MANETs) are networks made of mobile nodes
that self-configure and collaborate to forward packets among each other
without the benefit of an access point.  These networks are especially
important when infrastructure is unavailable (e.g., unplanned
networks, in remote areas, or after a disaster), or expensive.  In
such networks, each node must play the role of a router as well as a
station.

 Most existing MANET research assumes idealized wireless propagation:
nodes have a fixed transmission range and all receivers within this
range receive a transmission correctly (assuming no collision occurs).
However, due to fading and multipath effects this assumption
deviates significantly from reality: wireless transmission can suffer
deep fading (drop in power level) with very small changes in location,
or in time.  Fades of 20dB (the signal dropping to 1\% of its ideal
value) are not rare~\cite{Rappaport}.  These properties have profound
implications on protocol performance and designs that make many of the
decisions taken under idealized assumptions invalid.  Wireless
transmission success ratio drops with the distance between sender and
receiver.  This behavior leads to several interactions within the MAC
layer and across layers, some of which are subtle.

The effect of fading on MANET behavior has been observed and
studied empirically;  Decouto et al show that,
contrary to idealized models, fading has considerable effect on
link state -- using shortest path as a measure of
path quality can therefore be misleading~\cite{decouto-03a}.  Moreover, the packet loss
rate varies from link to link (as a function of the distance and the
surrounding environment); the quality of a hop must be made visible to
the routing protocol to enable effective route selection.  To capture
this effect, they propose ETX: a link cost metric that is a function
of the forward and backward delivery ratio on a
link~\cite{decouto-03}.  Other link cost metrics have also been
proposed to allow fading sensitive route selection~\cite{adya-04}.
Draves et al conducted an experimental analysis of these metrics and
found that ETX performs best~\cite{draves-04}.  Integrating these
effects into a routing architecture has been studied as well.  Woo et
al explored a routing architecture for managing wireless propagation
vagaries in a sensor network environment~\cite{woo-03}.  One of their
main conclusions is that using ETX as a route metric provides stable
routing performance.  However, the static sensor network environment
allows specialized routing and link estimation solutions that are
difficult to generalize to a MANET environment.  Draves et al propose
Link Quality Source Routing (LQSR), an extension of the Dynamic Source
Routing Protocol to address the problems that arise due to
fading~\cite{draves-04}.  In addition to allowing path selection
based on link quality, LQSR has several additional interesting
features such as continuous monitoring of path quality and higher
route stability.

Other effects of fading have also been observed and are reviewed
throughout this paper.  The majority of the work is experimental.
More specifically, problems are observed in real testbeds and
solutions are proposed to address them.  As a result most of the work
is focused on routing implications.  In this paper, we take a
complementary bottom up approach to this problem.  We start with an
analysis of the problem at the MAC layer.  As a result, we capture
some problems that occur in the MAC layer itself as well as problems
that affect the upper layers.  The bottom up approach provides a more
systemic and comprehensive evaluation of the effect of
fading on protocol performance than experimental testbed analysis.
The contributions of the paper also include the identification of
new problems that arise due to fading and the development
of initial solutions for them.

We first characterize the effect of fading on the performance MANET
protocols analytically and using simulation using a slow fading
propagation model.  One of our contributions is to show that Collision Avoidance (Request to Send (RTS)/ and Clear to Send (CTS) packets of IEEE 802.11)
is harmful to performance under fading and not that beneficial for
collision avoidance.  Moreover, we identify an unfairness problem that
occurs due to unbalanced backoffs that occur due to fading losses.
In addition, we outline the implications of this behavior on the
routing layer and propose preliminary solutions to the problems.

The remainder of this paper is organized as follows.
Section~\ref{background} explains some background
information. Section~\ref{sec:mac} analyzes the basic effect of
fading on MAC layer performance. Section~\ref{sec:noRTS} makes the case for
eliminating Collision Avoidance from MAC protocols.  Section~\ref{sec:backoff}
outlines the unfairness and inefficiency problems occurring due to backoff under fading.  In Section~\ref{sec:routing} we present high level descriptions of additional problems that occur
at upper layers.  Finally Section~\ref{sec:conclude} presents some concluding
remarks.

\section{Background}~\label{background}

\subsection{Wireless Propagation}

In a typical terrestrial environment, the transmitted signal reflects
off of surrounding objects, refracts when travelling through obstacles
and suffers diffraction due to sharp edges.  Moreover, doppler shifts
occur due to moving objects.  As a result, many echos of the
transmitted signal are received with different delays and power-levels
depending on the path they took.  Together, these result in large
transient flucuations in the power level: this is known as
fading~\cite{Rappaport}.

The details of wireless propagation are beyond the scope of this
paper.  Briefly, fading of a wave can be explained by slow fading
(e.g., due to dominant shadowing objects) and fast fading (due
to numerous smaller objects)~\cite{Sklar-97}.  Slow fading occurs over
time periods generally longer than a packet length; thus, we may
consider a single transmission power value per packet.  Alternatively,
small scale fading, or fast fading, occurs within the packet.  Fast
fading model considers the effect of scattered wave (also called by
multi-path reception model).  The Rayleigh or Ricean statistical
distributions have been shown to capture this effect well.  Ricean is
used when there is a line of sight path between sender and receiver
and Rayleigh is used when there is none.

With advanced modulation technologies (specifically, spread
spectrum/CDMA) and specialized antennas (such as the RAKE receiver),
the effect of small scale fading can be almost eliminated in the RF
frequency bands used for wireless communication.  Thus, we focus only
on slow fading.

\subsection{Propagation Models}

There are three kinds of propagation models typically used in MANET
protocol simulation and analysis.  The simplest model is Free Space
model.  The energy is in inverse proportion of square of distance.
This model is too simple to apply to realistic terrestrial setting.
Therefore, a more realistic model (called Two Ray Ground), which
considers the reflection of signal against the ground as well as
directly propagated signal, is used.  At short distance, only the
directly propagated signal matters.  Thus, the path loss exponent
($\beta$), which determines the (exponentiated) rate of attenuation of
the signal with distance, is 2; i.e., the power drops with the square
of distance.  However, after a cross-over distance which depends on
the height of the antennas, both the direct component and the ground
reflected component combine to create a higher path loss factor
(typically assumed to be between 3 and 4).  The Two-Ray Ground Model
is also idealized; it does not consider the fading effects
described above.

We use a fading propagation model in this paper that statistically
models slow fading.  Although this model is known (e.g., it is
available in the NS-2~\cite{ns-2} and Qualnet
simulators~\cite{qualnet}), it is not typically used.  In this model,
the path loss is a random variable that has a log-normal distribution,
with a mean equal to the expected two-ray ground path loss.  More
specifically, the signal power consists of two parts: the mean power
and the fading effect. The mean received power at a communication
distance is the idealized power as a function of that distance is
calculated as follows,

\begin{equation}
[\frac{\overline{P_r(d)}}{P_r(d_0)}]_{dB} =  -10\beta log(\frac{d}{d_0})          
\label{eq4}
\end{equation}
where $d0$ is a reference distance that is a function of the antenna
height.  The path loss equation in formula dB is eq~\ref{eq4}.
$\beta$ is the ideal path loss exponent (i.e., without considering
fading).

The second component models the transient fading effect.  Received
power is adjusted by a log-normal random variable $X_{dB}=
(N(0,\sigma^2))$.  The fading is modeled as a gaussian distribution
with average 0, and standard deviation $sigma$.  The overall received
power ius expressed as,
\begin{equation}
[\frac{\overline{P_r(d)}}{P_r(d_0)}]_{dB} =  -10\beta log(\frac{d}{d_0})  + X_{dB}.
\label{eq5}
\end{equation}
Because random variable $X_{dB}$, the range no longer represents a
discrete threshold with ``in-range'' or ``out-of-range'' nodes.
Rather, it is now continuous: packets in the ideal range may be lost
and packets outside the ideal range may be received.  The probability
of correct reception decreases with distance accoding to
Eq.~\ref{eq5}.  While this model is available in simulators such as
NS-2 and QualNet, it is almost never used
in MANET network level research.  We use it as the basis for our
analysis.

Rather than focusing on the impact of $\sigma$ and $\beta$ on protocol
behavior, we pick representative values and use them to evaluate the
impact of fading on MANET protocols.  Clearly, the impact can be
amplified or lessened with different $\sigma$ and $\beta$ values.

\subsection{Fading Model Limitations}

The model we use is statistical and has several limitations.  Recent
observations have shown that transmission losses at the link layer are
not generally independent~\cite{aguayo-04}.  Moreover, a limitation of
the model we consider is that the PDF of the received power is
strictly a function of distance, making links symmetric.  However,
fading is known to cause asymmetric link qualities.  Since IEEE
802.11 requires symmetric communication, it is likely to suffer as a
result of asymetric link qualities.  The effect of Asymmetry on
routing behavior has been considered but we do not know of any
analysis of the microeffects that arise due to
asymmetry~\cite{marina-02}; this is a topic of future work.  There is
a need for developing more accurate propagation models that capture
these effects.  However, we believe the simple fading models we use
are sufficient to identify the effects and problems we discuss in this
paper: more refined models will certainly enable more accurate
characterization of their effect.

Another limitation of the model is that we do not consider the effect
of multi-rate MAC protocols.  Recent versions of IEEE 802.11 recognize
the effect of fading on link quality~\cite{ieee-802.11std}.  To combat
this effect, the use multi-rate transmission where the modulation
scheme used is adapted dynamically and at the packet level to match
the link quality (high rate when link quality is high and lower rate
when it is low)~\cite{sadeghi-02}.  While this mitigates the effect of
fading, it comes at a high cost because the interframe spacing,
backoff periods and the rate negotiation headers have to be
transmitted at the lowest rate~\cite{pagtzis-01}.  We focus on single
transmission rate MAC protocols. While multi-rate MACs help to
mitigate the effect of fading, low-cost, low-power radios continue
to be single rate~\cite{bluetooth,ieee802.15std,woo-01}.  These radios are
the most likely candidate for use in embedded and low power devices.

\section{Basic Effect of Fading on the MAC Protocol}~\label{sec:mac}

In this section, we analyze the effect of fading on the MAC layer.
The effects identified in this section are not intended to be
comprehensive.  The fading model used is idealized and the
scenarios are simple.  Other intricate interactions may arise in more
realistic environments.  Our aim is to emphasize that fading must
be treated as a first class problem in MANET MAC protocol design.
Furthermore, in the next Section, we argue that it must also be
carefully designed for in higher level protocols.

The NS-2 network simulator is used for all simulation experiments in
this paper~\cite{ns-2}.  We use a path loss component of 3 (typical
range is 2--6 in real environments).  For the log-normal fading
model we set $\sigma=4dB$ (recall that 20dB fades are not uncommon; a
standard deviation of 4dB is rather conservative).  The transmission
power is set such that the ideal range is 250 meters.

\begin{figure}
\centerline{\includegraphics[width=0.85\linewidth]{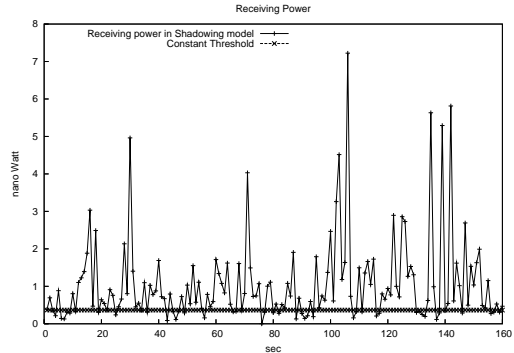}}
\caption{Unstable receiving power} 
\label{fig-recvpower220}
\end{figure}

 Figure~\ref{fig-recvpower220} shows a simulation trace of the
received power at distance 220 meters.  Note that when the receiving
power drops below the receiver sensitivity threshold, the transmission
is lost. Because we use a slow fading model, the same receive power is
assumed per packet.  

\subsection{Preliminaries}

In fading model, the received power depends on path loss $\beta$
and the log-normal component $X_{dB}$.  Eq (~\ref{eq5}) can be
simplified as the following: $\psi (p) = \tau(d) + X_{dB}$, where
$\psi(p)=[\frac{\overline{P_r(d)}}{P_r(d_0)}]_{dB} $ and
$\tau(d)=-10\beta log(\frac{d}{d_0}) $. Recall that $\psi(p)$ has
normal distribution $N( \tau(d) ,\sigma^2 )$.  Ignoring interference,
the probability of correct transmission is the probability of the
received power being higher than the receiver sensitivity threshold
($P_{th}$, which is a radio constant).  This can be directly computed
as P($\psi(p) \ge P_{th}$).  

\begin{figure}
\centerline{\includegraphics[width=0.85\linewidth]{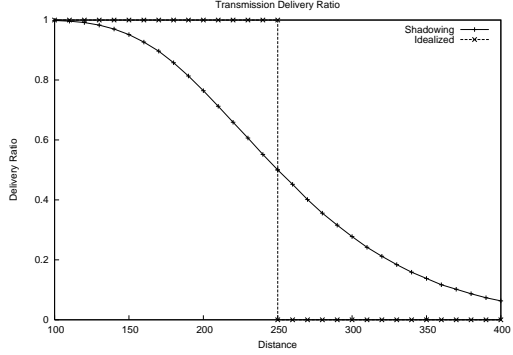}}
\caption{Delivery Ratio With Distance} 
\label{fig-deliveryratio}
\end{figure}

Figure ~\ref{fig-deliveryratio} shows the computed delivery ratio
vs. that of the idealized two ray ground model.  Clearly fading has
a large effect on the packet delivery ratio as the distance between
the sender and the receiver increases.  Although we show our results
as function of distance, they should be considered of in terms of
delivery ratio (which in our model has a one to one mapping per
Figure~\ref{fig-deliveryratio}).  In real testbeds, the packet
reception probability is influenced by the surrounding environment and
its PDF is not only a function of distance.

This observation confirms the need for routing protocols to be aware
of the link quality. The figure also shows the delivery ratio obtained
from simulation which not surprisingly, is very close to the
calculated ones.  Since the link level transmission ratio in the
simulation is being generated with the same distribution used in the
analysis, the simulation represents a Monte Carlo solution with
packets dropped according to the propagation model equation.

\subsection{Effect of Packet Retransmission}

Packet retransmission is used to increase reliability and to recover
both from transmission errors and collisions.  In this section we
analyze the effect of retransmission on packet delivery ratio and show
that under low delivery retransmission is in fact counterproductive.
Moreover, we analyze the effect of retransmission on packet delay.  We
do this using the retransmission algorithm of IEEE 802.11.

In IEEE 802.11, there is a Short Retry Limit (SRL) and Long Retry
Limit (LRL).  A transmission is classified as long or short based on
its length relative to a fixed threshold.  A transmission below the
threshold is counted against the SRL, while a transmission above it is
counted against both SRL and LRL.  SRL and LRL are 7 and 4
respectively.  

\subsubsection{Effect on Packet Delivery Ratio}

Successful delivery of a data packet requires successful delivery of
sequence of RTS, CTS, DATA and ACK packets.  If the data packet is
shorter than the threshold, 7 retransmissions are tried regardless of
where the failure is.  If it is longer than the threshold, failures
that occur in the DATA or ACK are counted against both limits (since a
long transmission of the data has occurred), but failures in RTS or
CTS count against SRL only.

The effect of retransmission on overall packet delivery in the
presence of fading (but ignoring collisions) can be derived as
follows.  In order for a transmission to succeed, RTS, CTS, DATA and
ACK should not fail.  Thus packet success ($p_s$)and failure ($p_f$)
probabilities on a given try are given by $p_s=p^4$ and $p_f=(1-p^4)$,
where p is the link level delivery ratio (Eq~\ref{eq5}).  If there is
a failure in the first try at either of these four transmissions, a
second retry occurs, and so on.  The success in the first four tries
can be computed (binomial experiment) as:
\begin{eqnarray}
P_{1to4}= p_s+p_f*p_s+p_f^2*p_s+p_f^3*p_s \nonumber \\ 
p_s(1+p_f+p_f^2+p_f^3)
\end{eqnarray}
A fifth transmission will only occur if at least one of the failures
did not count against LRL.  The probability of a short failure
(failure in RTS or CTS) can be computed as
\begin{equation}
p_{sf} = 1-p^2
\end{equation}
Similarly, the probability of a long retry failure is
\begin{equation}
p_{lf} = p^2(1-p^2)
\end{equation}
The probability of a success in the fifth transmission can be computed
as the probability of failure in all 4 first transmissions, with one
or more of the failures being short, and a success in the fifth try.
Specifically, 
\begin{equation}
P_5=(p_f^4-p_{lf}^4)*p_s
\end{equation}
Similar reasoning can be applied to compute $P_6$ and $P_7$ to give a
total packet delivery probability as follows.
\begin{eqnarray}
P_{packet}=p_s(1+p_f+p_f^2+p_f^3+ p_f^4-p_{lf}^4 \nonumber \\
+((4*p_{sf}^2*p_{lf}^3) + (p_f^4-p_{lf}^4  \nonumber\\
-4 p_{sf}*p_{lf}^3)*p_f)) + (16*(p_{sf}^3*p_{lf}^3))\nonumber \\
+(4*p_{sf}^3*p_{lf}+p_{sf}^4)*p_f^2))~\label{eqcbrpckcase}
\end{eqnarray}
\noindent
 For short length data packet, the packet delivery ratio can be
derived as the probability of success in the first 7 trials of a
binomial experiment with $p_s$ outcome probability.

\begin{figure}
\centerline{\includegraphics[width=0.85\linewidth]{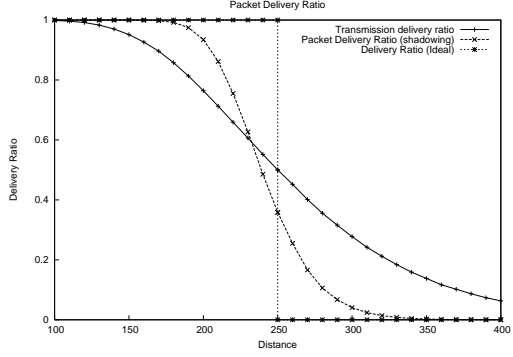}}
\caption{Packet Delivery Ratio} 
\label{fig-cbrdeliveryratiobycalc}
\end{figure}
Figure ~\ref{fig-cbrdeliveryratiobycalc} shows the probability of data
packet delivery.  Perhaps surprisingly, the long length data delivery
ratio is almost identical to the short length one; this indicates that
the high number of transmissions may not be needed (we note that this
conclusion may change when collisions are considered).  When the
delivery ratio is above 0.6, CBR packet delivery ratio is better than
link level delivery ratio.  However, {\bf below 0.6, the packet delivery
ratio is worse than the link delivery ratio despite retransmission.}
This effect occurs because correct retransmission requires correct
reception of 4 packets (RTS/CTS/DATA/ACK), while link level success
requires correct reception of only 1.  The retransmissions are not
sufficient to overcome this disadvantage at low link level delivery
ratios.

\subsection{Effect on Delay}

Fading also affects the delay of packet delivery; multiple retries
are needed, with an exponentially increasing backoff between them.  We
now develop a simple analytical model for the delay.  Developing a
closed form solution for packet delay in the presence of the long and
short retry limits requires enumerating possible sequences of failures
(a few hundred cases).  Instead, we develop a solution for the
expected value of a backoff period as a function of the link level
delivery ratio.

We note that the backoff is reset to the minimum value (31 slots)
whenever a CTS or ACK packet is received, and doubled whenever they
fail to be received (with a cap of 1023 slots).  The backoff period is
selected randomly between 0 and the current window size.  A CTS (ACK)
is received only if the RTS (Data) packet is received.  The
probability of two consecutive correct transmissions (ignoring
interference) is $p^2$.  The expected value of a backoff period can be
computed as follows
\begin{eqnarray}
E(Backoff) = \frac{p^2}{2} \cdot ( 31 + 63 *(1-p^2)  \nonumber  \\
+ 127*(1-p^2)^2 + 255 \cdot (1-p^2)^3  \nonumber  \\
+ 511 \cdot (1-p^2)^4) + \frac{1023}{2} \cdot (1-p^2(1  \nonumber  \\
+ (1-p^2) + (1-p^2)^2 +(1-p^2)^3 +(1-p^2)^4))
\end{eqnarray}
\noindent

\begin{figure}
\centerline{\includegraphics[width=0.85\linewidth]{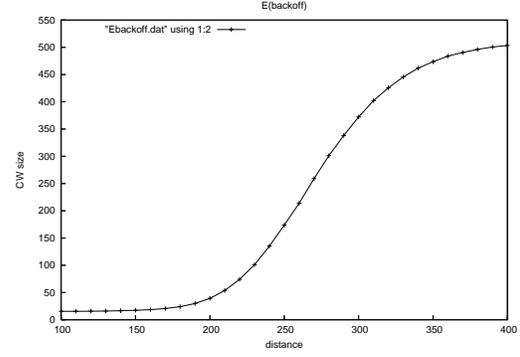}}
\caption{Expected Value of the Backoff} 
\label{Ebackoff}
\end{figure}

\begin{figure}[ht]
\centerline{\includegraphics[width=0.85\linewidth]{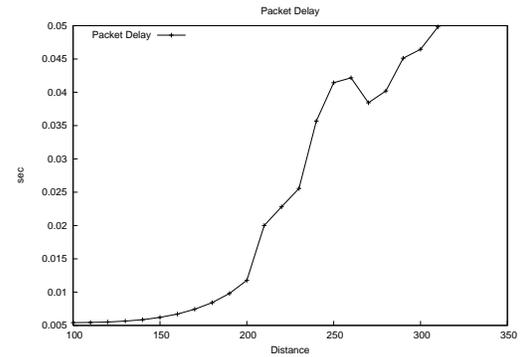}}
\caption{One-hop Packet Delay}
\label{delay1hop}
\end{figure}

This expected value is shown in Figure~\ref{Ebackoff}.  The average
backoff period is significantly higher than minimum even at reasonably
stable links.  The backoff period increases exponentially but is
capped at the maximum period resulting in the behavior shown.  Note
that the average packet delay will increase faster than this ratio
since the number of required retransmissions will increase as the
delivery ratio drops (incurring multiple backoff periods in addition
to the retransmissions).  The one-hop packet delay obtained by
simulation is shown in Figure~\ref{delay1hop}.  The delay increases
quickly with the as the quality of the link drops.  

In the following sections, we discuss some important implications that
follow from this analysis.  First, we make the case that RTS/CTS is
harmful to performance under fading.  Second, we identify an
unfairness problem that arises due to fading.  Finally, we discuss
additional effects that arise due to fading.

\section{RTS/CTS Considered Harmful}~\label{sec:noRTS}

As was observed in Figure~\ref{fig-cbrdeliveryratiobycalc}, the packet
delivery ratio drops below the transmission delivery ratio when the
link state is poor (below 60\% transmission delivery ratio).  The
result is explained by the fact that 4 transmissions must succeed for
a packet to be delivered (RTS,CTS, Data and ACK).  At low transmission
success probability it becomes highly improbable for 4 consecutive
transmissions to succeed.

An alternative approach (which is supported by IEEE 802.11 for short
packets) is to rely on just CSMA without RTS/CTS and use
acknowledgments to recover from errors and collisions.  We first
analyze this approach assuming no collisions and then revisit to
discuss the effect of collisions.  In this approach, only two
consecutive transmissions (Data and ACK) have to be received
correctly; this has a much higher probability of success than 4
consecutive transmissions.  In this case, only two transmissions need
be received correctly (DATA and ACK).  In this section, we make the
argument that removing RTS/CTS is beneficial for performance.  We make
this argument in two parts: first we show that removing RTS/CTS
significantly improves packet delivery in the presence of losses, and
then show that the effect of eliminating RTS/CTS on reducing
collisions is not large.

\subsection{MAC-layer Analysis without RTS/CTS}

  The probability of correct reception with a retransmit limit of 4
when RTS/CTS is not used can be obtained as a 4 step binomial
experiment as follows:
\begin{equation}
P_{packet}=p^2(1+(1-p^2)+(1-p^2)^2+(1-p^2)^3)
\end{equation}

\begin{figure}
\centerline{\includegraphics[width=0.85\linewidth]{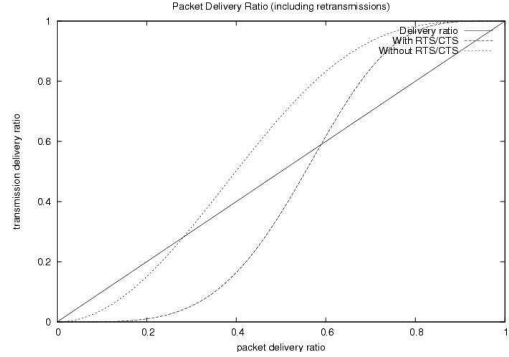}}
\caption{Packet Delivery with/without RTS/CTS} 
\label{no-rts-delivery}
\end{figure}

\begin{figure}
\centerline{\includegraphics[width=0.85\linewidth]{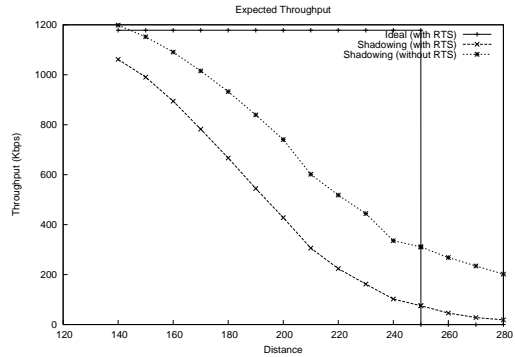}}
\caption{Link Capacity} 
\label{capacity}
\end{figure}

Packet delivery ratio is significantly improved when the RTS/CTS
handshake is omitted.  This effect can be seen in
Figure~\ref{no-rts-delivery}.  The average packet delay is reduced if
RTS/CTS are omitted.  This can be proven using the observation that
the expected delay of a transmission assuming all RTS/CTS are
successful and ignoring the cost of RTS/CTS is equivalent to the
expected delay of the approach that does not use RTS/CTS.  Thus, using
RTS/CTS, we have to add the cost of the RTS/CTS as well as failed
RTS/CTS and the associated increase in backoffs.
Figure~\ref{capacity} shows the obtained capacity with and without
RTS/CTS for a single hop -- clearly, there is a large advantage for
the case without RTS/CTS.  Similar positive results hold for the delay
(not shown).

\subsection{Effect on Collisions}

A potential drawback of this approach is that the advantages of
collision avoidance (RTS and CTS) in reducing collisions and collision
cost are lost. We argue that RTS/CTS is only of limited success in eliminating
collisions and that the complimentary Carrier Sense Multiple Access
(CSMA) is more effective in reducing collisions.  Some collisions
occur when two senders sense the medium to be idle and transmit
concurrently; such collisions cannot be prevented by either approach.
However, the vulnerability period to such collisions is low and their
effect is not likely to dominate.  One advantage that RTS/CTS offer to
such collisions is that contention is carried out with small control
packets, whereas a full data packet is lost if only CSMA is used.

\begin{figure}
\centerline{\epsfig{file=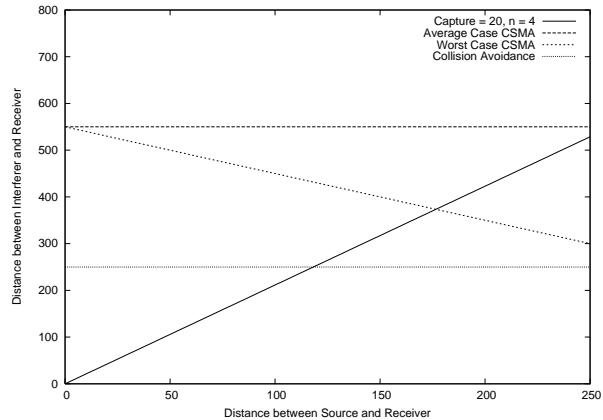,width=0.7\linewidth,angle=270,silent=}}
\caption{CSMA vs. Collision Avoidance} 
\label{csma}
\end{figure}

Collisions occur in a wireless environment if the ratio of the
received packet power to the interfering power (other transmissions
and noise) fall below a threshold called the {\em capture threshold}.
Its well known that collision avoidance is not sufficient to prevent
collisions: a node that is close enough to the receiver to interfere
but not close enough to receive the CTS packet can possibly cause a
collision.  In Figure~\ref{csma}, we consider a situation where
a source (S) is communicating with a receiver (R) in the presence of
an interfering node (I).  The x-axis shows the distance between the S
and R, while the y-axis shows that between I and R.  Points on the
figure above the capture lines indicate that capture does happen (the
interferer is too far) while those below indicate that a collision
happens.  The capture value represents the capture threshold, while
$n$ represents the path loss factor (the signal decays with
$\frac{1}{r}^n$ where $r$ is the distance from the transmitter).  The
capture line plots the interferer distance that would
cause $\frac{P_{sender}}{P_{interferer}}=Capture$.  Note
that this approximate analysis assumes only a single interferer,
ignores the effect of noise, and assumes ideal propagation.

Essentially to control collisions all potential interferers (below the
capture line on Figure~\ref{csma}) should be blocked.  Collision
Avoidance (RTS/CTS) can only block interferers in reception range of
the receiver (to receive the CTS).  Thus, many potential interferers
are not blocked with the RTS/CTS mechanism.  This is the area under
the capture line but above the 250 meter line.

The main mechanism for avoiding collisions in commercial wireless
cards is an aggressive Carrier Sense (with low sense threshold).  This
aggressive threshold is used both to attempt for interferers out of
receivers range and for the fact that carrier sense occurs at the
sender, but collisions occur at the receiver.  For a WaveLAN
card~\cite{wavelan-manual} with a nominal transmission range of 250
meters, the carrier sense threshold is set such that transmitters in a
circle of radius 2.2 times the transmission range around S are forced
to stay idle (assuming a path loss factor of 4).  Having this
aggressive threshold reduces collisions (it does not completely
eliminate them), but increases the exposed terminal problem.

CSMA's effectiveness in blocking interferers is a superset of that of
Collision Avoidance: all collisions that can be eliminated by
collision avoidance can also be eliminated by CSMA.  Furthermore, CSMA
is able to prevent many more collisions not preventable by RTS/CTS.
Since CSMA is applied at the sender but the interference happens at
the receiver, its effectiveness depends on the location of the
interferer from the source.  More specifically, for carrier sense to
occur successfully, the signal power of the interferer at the sender
should be above the carrier sense threshold.  In the average case
(when the interferer to source distance is equal to the receiver to
source distance; that is, the interferer location is not biased either
closer to the sender or further away from it), CSMA is able to prevent
all possible collisions (the average-case line on Figure~\ref{csma}).
The worst case for CSMA occurs is when the interferer is on the side
of the receiver away from the sender (this is the worst-case line on
Figure~\ref{csma}).  In this case, the interferer's distance to the
sender is maximized for a given interferer-receiver distance.  Even in
the worst case, CSMA prevents all collisions catchable by CA -- recall
that all the cases under the line are prevented for each mechanism.
Note that in the worst case, some collisions are not preventable.

Thus, we perceive no benefit for RTS/CTS in reducing collisions.
Their only benefit is in reducing the cost of collisions for
collisions that occur due to concurrent sensing of an idle channel; in
this case, collisions occur on the small RTS/CTS control packets
instead of full length data packets.  This only works for interferers
in range to receive a CTS -- this is a relatively small area of
possible interferers as seen on the Figure.  Moreover, in the case of multiple
interferers, their combined interference power may cause a collision.  This
helps CSMA and hurts CA: CA requires that each of the intereferers be in
reception range, while CSMA naturally takes their effect in since carrier
sense measures their combined effect.  Thus, given the
dramatic degradation CA causes on packet delivery ratio in the presence of
fading and the small benefit it provides to collisions 
it is beneficial to eliminate them.  

We note that this conclusion is only dependent on the carrier
sense threshold.  If the path loss component is different, or if the
send power is different, this affects both the carrier sense and the
collision avoidance.  Furthermore, if the capture threshold is
different, this has the same effect on both CSMA and CA (the slope of
the capture line changes, but not the other lines).

\section{Backoff Implications -- Inefficiency and Unfairness}~\label{sec:backoff}

The second major implication from the basic MAC analysis in
Section~\ref{sec:mac} concerns the backoff mechanism.  The backoff
mechanism in contention based MAC protocols (such as 802.11 DCF mode
used for MANETs) is intended to regulate the offered load to the
shared medium.  The underlying assumption is that all packet losses
are due to collisions.  While this assumption is true in wired shared
media where errors are exceptionally rare, it is not true in wireless
environments.  As a result of losses due to fading, the backoff
timer is increased.  This leads to two important side effects: (1)
Inefficiency in using the medium: backoffs occur even without
collisions, leading to nodes backing off excessively
causing unnecessary channel idle time;
and (2) unfairness: because the expected backoff period increases with
the of transmission losses, links that experience losses have a larger
average backoff than those that do.  IEEE 802.11 is known to be
susceptible to short-term transient unfairness even under idealized
propagation assumptions~\cite{vaidya-00}.  Under fading
we show that steady-state unfairness can occur.

The inefficiency issue was already alluded to in the discussion of the
delay and effective throughput.  Backoff increasingly contributes to
the delay as the transmission delivery ratio drops: as a larger number
of retries is needed, we have the multiplicative effect of a larger
number of backoff periods and longer average backoffs.  The problem
occurs due to the implicit assumption that losses are due to
contention.  In the presence of fading, this is often not the case
causing inappropriate backoff.  Finer discrimination between
contention losses (collisions) and transmission losses (fading) are
needed.

The unfairness problem occurs due to the imbalance in backoff
durations.  On average, links with lower transmission success ratios
will have a higher backoff period (Figure~\ref{Ebackoff}).  As a
result, competition for the link is no longer fair -- links with
higher transmission delivery ratio have a higher probability of
capturing the link.  IEEE 802.11 already suffers from short term
unfairness~\cite{vaidya-00}.  However, the unfairness problem
identified here is sustained.  Addressing this problem also requires
backoff algorithms that can discriminate between collisions and
transmission losses.

\begin{figure}
\centerline{\includegraphics[width=0.85\linewidth]{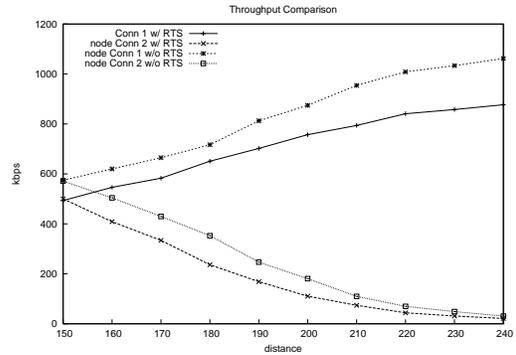}}
\caption{Unfair Throughput} 
\label{unfair3}
\end{figure}

To illustrate this problem, we simulate two single hop connections
whose sources are in range with each other, but whose receivers are
not (to isolate the effect of a single contention point).  The
distance (and therefore the transmission delivery ratio) of one
connection was fixed at 150 meters (Connection 1) and the other varied
(Connection 2).  Each connection generates CBR traffic at a rate that
would use all the available bandwidth if there is no contention.
Figure~\ref{unfair3} shows the raw throughput obtained by the two
connections with and without RTS/CTS.  The first observation is that
the no RTS/CTS version is able to
obtain higher throughput even in the presence of contention.  The
unfairness problem can be seen on the diagram as the first connection
gets an increasingly higher portion of the available bandwidth at the
expense of the weaker connection.

\begin{figure}
\centerline{\includegraphics[width=0.85\linewidth]{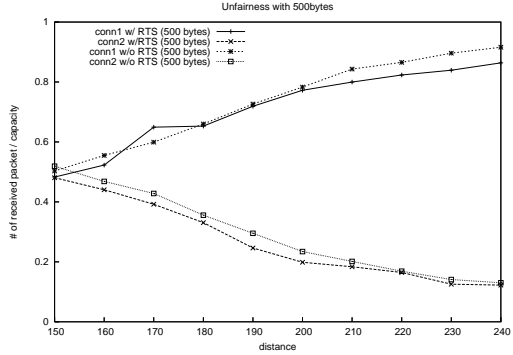}}
\caption{Unfairness with Packet size 500 bytes} 
\label{unfair1}
\end{figure}

\begin{figure}
\centerline{\includegraphics[width=0.85\linewidth]{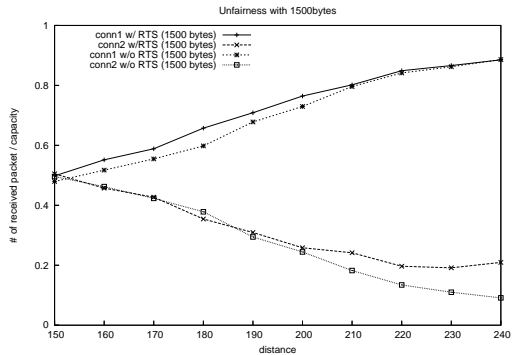}}
\caption{Unfairness with Packet size 1500 bytes} 
\label{unfair2}
\end{figure}

Its difficult to assess the degree of unfairness since the expected
throughput of the hop goes down with the increased hop delay as the
connection becomes weaker.  To normalize this effect, we use a metric
of the percentage of delivered packets as a ratio of the maximum
deliverable packets on the connection if no interference was present.
In a fair implementation, the two connections would be able to get an
equivalent percentage of their maximum throughput.
Figures~\ref{unfair1} and ~\ref{unfair2} illustrate the problem for 500
byte and 1500 byte packets respectively.  Clearly, the strong
connection dominates for both scenarios.  When the two distances are
equal, the connections share the bandwidth almost equally since the
backoffs are not biased in favor of either connection.  The unfairness
problem is clear as the connection using the strong link quickly
dominates the weaker one.  This is true for both packet sizes studied
and with and without RTS/CTS.

\begin{figure}
\centerline{\includegraphics[width=0.85\linewidth]{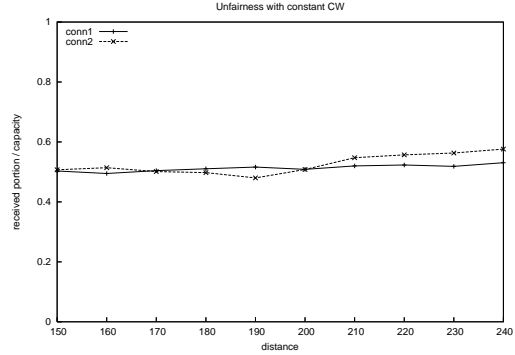}}
\caption{Backoff Disabled} 
\label{fair}
\end{figure}

To verify that the unfairness is due to backoff, we disabled backoff
in the above scenario since the vast majority of losses were due to
fading.  In this case, the two connections were able to get fair
access to the medium across all distances (Figure~\ref{fair}).  This
naive algorithm assumes that all losses are due to fading.
Clearly, this is not a feasible solution, but it highlights the
importance for discrimination between fading and contention losses.
For example, possible discriminators such as channel utilization as a
measure of contention or the use of physical level detection of
collisions at the receiver with feedback to the sender are likely to
provide effective discriminators for controlling the MAC backoff
algorithm.  This is a topic of future work.

\section{Effects on Upper Layers}~\label{sec:routing}

A number of problems due to fading have been encountered in real
testbeds and several solutions have been developed.  The most heavily
studied problem is the issue of link quality and exposing that to the
routing protocol to enable it to evaluate path
quality~\cite{decouto-03a,aguayo-04}.  Moreover, the problem of
discovering low quality hops that may become visible due to fading
is well known~\cite{chin-02,maltz-00}.  The problem occurs if the protocol
attempts to discover only high quality links -- lower quality ones are
still occasionally visible to route discovery.  One of the proposed
solutions is to use MAC filtering to assess the quality of the
link before accepting it~\cite{chin-02,maltz-00,dube-97,goff-01}; this approach
improves performance but does not eliminate the problem.   We were able to
reproduce these problems in our simulations.   
In this section, we outline additional problems and interactions that arise 
at the upper protocol layers due to fading.  We are pursuing solutions to several of these problems.

\subsection{Link and Route Qualities}
Exposing link quality to the routing protocol is an important step
towards effective evaluation of route qualities.  However, we believe
that current link quality metrics and their combination into a route
costs are not representative of actual behavior.  Packet delivery
ratio based metrics such as ETX appear to perform better than delay
metrics~\cite{draves-04,woo-03}.  The route quality is then obtained by
adding up the costs of the individual links.  However, both link delay
and throughput are not linear functions of packet delivery.  Simply
adding the individual links does not provide an accurate estimate of
the delay or expected throughput of the connection.

A more subtle effect occurs at the connection level as well.  Due to
the unfairness problem, the problem of self-contention among hops of a
single connection exhibits markedly different performance when
fading is considered.  Specifically, under idealized assumptions,
there is a bias towards hops closer to the source since these hops
supply packets.  Thus, later hops cannot be unfair against earlier
hops -- the best they can hope for is to match
them~\cite{xu-02}.  When fading is considered, the
effect of this problem is much more pronounced.  More specifically,
because a better quality link can dominate the available bandwidth
when contending with a lower quality one, the following effect is
observed.  If a strong connection precedes a weaker one in a multi-hop
chain, the strong connection will dominate the weaker one.  Thus,
packets would be sent to the source of the weaker connection, and get
dropped there because it is unable to get a share of the bandwidth to
send the packets it received.

If the weaker link comes first, it has a regulating effect on the
stronger one -- the stronger link is limited in terms of supply
packets to what the weaker link can provide it.  Thus, the quality of
the route is not simply the sum of the qualities of the links -- the
order of the links has to be considered as well.  Alternative,
sophisticated packet scheduling techniques can be used to bypass this
problem.  We believe that addressing the root problem of MAC level
unfairness will provide a better solution.  

\begin{figure}
\centerline{\epsfig{file=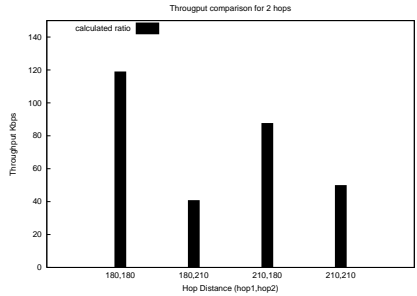,width=0.7\linewidth,silent=}}
\caption{Effect of Hop Order}
\label{order}
\end{figure}

The effect of the link
order on chain performance can be seen in Figure~\ref{order}.  In this experiment, the throughput of a two hop connection is tracked.  The figure shows 4 situations depending on the quality of the two links.  Most interesting is the performance of the middle scenarios.  A
connection with a weak link followed by a strong one performance much
better than a connection with the strong link first.  In fact, when the strong link comes first, the throughput is {\em worse} than a connection with 
two weak links (the fourth scenario).  
Our future research regarding this problem targets coming up with
generalized route quality estimates based on link qualities and order.  

\subsection{Effect of Spurious Route Errors}

The dual of discovering unwanted low quality hops is mistakenly
thinking that good quality ones are no longer there.  Some routing
protocols such as DSR~\cite{johnson-01} assume that a failure to send
a packet is due to mobility.  Again, this is due to idealized
propagation assumptions causing packet losses to be almost always due
to mobility.  However, packet failures due to repeated collisions have
been shown to cause route errors under heavy loads as
well~\cite{xu-02}.  Route errors can have a major effect on
performance -- leading to expensive route searches and connection
interruption while route discovery is accomplished.

Under fading, even in good quality links occasional packet failures
do arise.  For example, even when the transmission delivery ratio is
90\%, there is approximately a 2\% chance of a packet failure with 4
retries (as would happen with no RTS/CTS).  This would lead to route
error getting generated every 50 packets; the effect is even worse
when one considers a multi-hop connection -- each data packet has a
chance of causing a route error at every hop.

The Link Quality Source Routing (LQSR) protocol provides a framework
for managing this and similar problems~\cite{draves-04}.  LQSR
constantly monitors path quality by providing feedback to the sender
potentially with every packet delivered.  While LQSR still generates the
Route Error Packet when delivery failure occurs, it does not interpret
such a packet as a loss of route at the source -- simply, the source 
penalizes the link (increasing its cost estimate).  

Such approaches are needed to filter out transient losses due to
fading from real loss of route.  However, a large amount of
overhead can be generated due to these route error packets.  We have
developed an approach for locally discriminating between fading
losses and loss of connection due to mobility.  Essentially, an
exponential average of the delivery ratio along an active link is
tracked.  When packet drops occur, they generate route errors only if
the stable estimate of the link quality is low (indicating possibility
of disconnection).

\subsection{Effect on MAC broadcasts and Route Discovery}

\begin{figure} 
\centerline{\epsfig{file=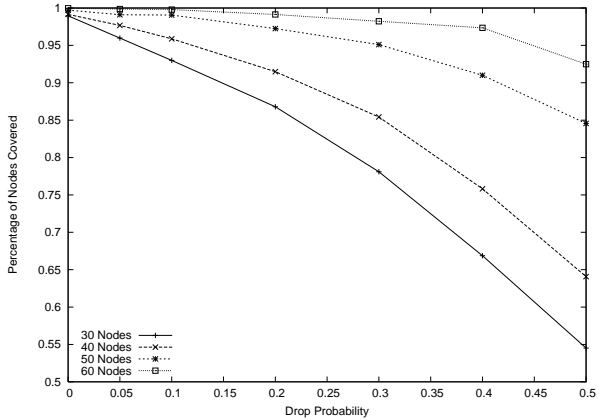,width=0.7\linewidth,angle=270,silent=}}
\caption{Effect on Network Wide Broadcast}
\label{paul-fig}
\end{figure}

Fading has especially high impact on broadcast operations.  Unlike
unicast packets, MAC level broadcast is not acknowledged -- if a
broadcast is loss, the loss is undetected and no retransmissions
occur.  As a result, broadcast transmissions are only delivered at the
link level transmission ratio (see Figure~\ref{fig-cbrdeliveryratiobycalc}), making
them especially vulnerable to fading.  This has important
implications on all protocols that use MAC broadcast such as flooding
and group communication operations.  

Figure~\ref{paul-fig} shows the effect of this problem on flooding
coverage as the network density increases.  In this scenario, a number of
nodes are deployed randomly in a 1000x1000 meter area; the number of nodes is varied to increase the density.  We
simulated the effect of fading using probabilistic packet drop to
have control on the loss rate.  Clearly, the coverage of flooding
suffers as the quality of the links drops.  This is especially true
for sparse networks (and for sparse areas of networks).  Moreover, since many optimized Network-Wide
Broadcast operations reduce the redundancy in flooding, they end up
becoming more vulnerable to fading losses.  We characterize this
problem and investigate solutions to it in another paper submitted to
MobiHoc. 

Because of fading packets may fail multiple consecutive
retransmissions and be dropped because the retransmission limit is
reached.  Many routing protocols use MAC level transmission failure as
an indicator that the link is no longer available (due to mobility).
It is important to have more effective discrimination between packet
losses due to fading and those due to mobility.

\section{Conclusions}~\label{sec:conclude}

In this paper, we presented the result of a bottom-up investigation of
the effect of fading on MANET protocol performance.  We first did
an analysis of the effect of fading on the MAC layer and showed
that packet delivery ratio, packet delay and effective throughput all
suffer as a result of fading.  

Based on this analysis, we made the case the Collision Avoidance (the
RTS/CTS mechanism) is harmful for performance.  This case has two
sides: under fading, it become significantly harder to deliver
packets if 4 successive transmissions must be delivered correctly
(RTS/CTS/DATA/ACK) vs. only 2 transmissions if RTS/CTS is not used.
In addition, we showed that RTS/CTS has minor benefit in preventing
collisions.  More specifically, with aggressive carrier sense (as is
commonly used in commercial radios), CSMA can prevent all collisions
that are also preventable by CA (and in fact, considerably more).  The
benefit of RTS/CTS is then isolated to making the cost of contention
for situations where an interferer is in range with either the sender
or the receiver less (due to contention using the smaller RTS
packets).  Overall, we believe that the large benefit in performance
due to eliminating RTS/CTS is not recuperated by the small benefit to
the cost of collisions for a subset of the collisions.

Another primary contribution of this paper is to identify an
unfairness problem that arises due to fading.  More specifically,
the MAC layer backoff algorithm presumes that all losses are due to
contention.  Thus, losses that occur due to fading increase the
backoff even when no contention exists.  Thus, weaker links are at a
disadvantage to stronger links because they end up backing-off more
frequently.  We showed that this problem can be severe even when the
difference in the quality between the links is minor.  In the long
term, we believe that the root of this problem must be attacked: the
backoff algorithm should be able to discriminate between contention
and transmission losses.  We are investigating techniques to estimate
the contention to avoid the inefficiency and unfairness problems that
arise due to it.  In addition, we outlined the effect of fading on
broadcast and multicast operations.

We also presented an overview of additional problems that arise at the
upper layers.  Some of these problems are known, but the solutions to
them are ad hoc in nature and not systemic (for example, the problem
of discovering low quality links; also the related problem of dropping
good quality ones when a packet fails).  One of the problems we
outlined is an artifact of the unfairness problem at the MAC layer.
As a result of this problem, path quality is sensitive to the order of
hops in a multi-hop connection.  More specifically, if a strong link
is closer to the source than a weaker one, it ends up dominating it
due to the unfairness problem: packets would constantly be delivered
to the source of the weaker hop only to be dropped there.  However, if
the weaker link comes first, the stronger link cannot dominate because
its packet supply comes from the weaker link.  There is a need for a
generalized approach for estimating link quality that takes such
behavior into consideration.

The fading model we use has several limitations.  It treats links
symmetrically and assumes that losses are independent; both these
assumptions are not valid in real testbeds.  Nevertheless, we believe
that the problems that are isolated are real but the magnitude of
their effect may be different when a more accurate model is used.
Moreover, we did not consider the effect of multi-rate MAC
protocols~\cite{sadeghi-02,seok-multirate}.  Upgrading the propagation
model to better reflect wireless propagation and generalizing our
analysis is also a topic of future work.

\bibliography{local}
\bibliographystyle{IEEE}
\end{document}